\documentclass[aps,prl,reprint]{revtex4-2}
\usepackage{amsmath,amssymb,graphicx,numprint,times,adjustbox, multirow,dcolumn,mathtools,xspace, verbatim}
\usepackage[usenames,dvipsnames,svgnames,table]{xcolor}
\definecolor{prlblue}{rgb}{0.18,0.18,0.573}
\usepackage[colorlinks=true, urlcolor=prlblue, linkcolor=prlblue, citecolor=prlblue, pdftex]{hyperref}
\usepackage{orcidlink}

\npthousandsep{\,}
\npdecimalsign{.}

\def\<{\langle}
\def\>{\rangle}

\newcommand{\nn}{\nonumber}
\renewcommand{\d}{\partial}
\newcommand{\cdag}[1]{\hat{c}_{#1}^\dag}
\newcommand{\cd}[1]{\hat{c}_{#1}^{\phantom{\dag}}}
\newcommand{\nop}[1]{\hat{n}_{#1}}
\newcommand{\pdag}[1]{\psi_{#1}^\dag}
\newcommand{\pd}[1]{\psi_{#1}^{\phantom{\dag}}}

\newcommand{\chidof}{\ensuremath{\chi^2/\text{d.o.f.}}\xspace}
\newcommand{\Lmin}{\ensuremath{L_{\text{min}}}\xspace}

\newcolumntype{L}{>{$}l<{$}} % math-mode version of "l" column type
\newcolumntype{u}{D{.}{.}{8}}
\newcolumntype{w}{D{.}{.}{10}}
\newcolumntype{q}{D{.}{.}{4}}

\newcommand{\prlsection}[1]{{\it #1.}---}

\newcommand{\switchtoletter}[1]{
\setcounter{equation}{0}
\renewcommand{\theHequation}{#1\arabic{equation}}
\renewcommand{\theequation}{#1\arabic{equation}}

\setcounter{figure}{0}
\renewcommand{\theHfigure}{#1\arabic{figure}}
\renewcommand{\thefigure}{#1\arabic{figure}}

\setcounter{table}{0}
\renewcommand{\theHtable}{#1.\Roman{table}}
\renewcommand{\thetable}{#1.\Roman{table}}
}
\newcounter{appcounter}
\newcommand{\prlappendix}[1]{
\addtocounter{appcounter}{1}
\switchtoletter{\Alph{appcounter}}
{\it Appendix \Alph{appcounter}: #1}---}

\newcommand{\PRL}{Phys.~Rev.~Lett.}

\newcommand{\PRX}{Phys.~Rev. X}

\newcommand{\JHEP}{J.~High Energy Phys.}

\newcommand{\JPCM}{J.~Phys.: Condens.~Matter}

\newcommand{\physrep}{Phys.~Rep.}

\newcommand{\NPB}{Nucl.~Phys.~B}
\newcommand{\JSF}{J.~Stat.~Phys.}

\newcommand{\RMP}{Rev.~Mod.~Phys}

\begin{document}

\title{Extraordinary transition at the edge of a correlated topological insulator}
\author{\firstname{Francesco} \surname{Parisen Toldin}\,\orcidlink{0000-0002-1884-9067}}
\email{parisentoldin@physik.rwth-aachen.de}
\affiliation{\mbox{Institute for Theoretical Solid State Physics, RWTH Aachen University, Otto-Blumenthal-Str. 26, 52074 Aachen, Germany}}
\author{\firstname{Fakher F.} \surname{Assaad}\,\orcidlink{0000-0002-3302-9243}}
\email{assaad@physik.uni-wuerzburg.de}
\affiliation{Institut f\"ur Theoretische Physik und Astrophysik and W\"urzburg-Dresden Cluster of Excellence ct.qmat, Universit\"at W\"urzburg, 97074 W\"urzburg, Germany}
\author{\firstname{Max~A.} \surname{Metlitski}\,\orcidlink{0000-0003-2587-0320}}
\email{mmetlits@mit.edu}
\affiliation{Department of Physics, Massachusetts Institute of Technology, Cambridge, Massachusetts 02139, USA}

% Abbreviations
% 3D: three-dimensional
% BCs: boundary conditions
% CFT: conformal field theory
% FSS: finite-size scaling
% LL: Luttinger liquid
% QMC: quantum Monte Carlo
% RG: renormalization group
% SM: Supplemental Material
% UC: universality class

\begin{abstract}
  The interplay of topology  and correlations  defines a new  playground to study boundary criticality in quantum systems. We employ large scale  auxiliary field quantum Monte Carlo simulations
  to study a two-dimensional Kane-Mele-Hubbard model on the honeycomb lattice with zig-zag edges and the Hubbard U-term tuned to the three-dimensional XY bulk critical point.  Upon varying the Hubbard-U term on the edge we observe a boundary phase transition from an ordinary phase with a helical Luttinger liquid edge decoupled from the critical bulk to an extraordinary-log phase  characterized by a logarithmically diverging spin stiffness.
  We find that the spectral functions exhibit distinct features in the two phases giving potential experimental signatures.
\end{abstract}

\maketitle

\prlsection{Introduction}
In recent years, there has been much interest in the study of extended operators, boundaries and defects. It is well known that gapped (topological) phases can host protected boundary and defect states, whose theory is by now highly developed.\cite{HK-10, WenCoho, LevinUngap, SenthilSPTRev,CatLectures} For critical bulk states the  understanding of boundaries and defects, while still far from complete, has been experiencing rapid progress, particularly when the bulk is described by conformal field theory (CFT).\cite{MO-95,LRVR-13,GLMR-15,BGLM-16,LM-17,MRZ-19,KP-20,BDPLVR-20, GGLVV-21, JensenOBannon, Zoharg, Zoharcusp, YifanFusion, kravchuk} A problem that has attracted much attention is whether topologically protected boundary states can exist in some form when the bulk of a quantum system is gapless.\cite{GV-12,SPV-17,PSV-18, Verresen-20,  VTJP-21, TVV-21} A natural setting to study this question is at a bulk quantum phase transition out of a topological phase that supports edge states. In this paper, we consider precisely such a scenario, and investigate via Quantum Monte Carlo (QMC) simulations the edge behavior of a two-dimensional quantum spin Hall insulator as its bulk undergoes an antiferromagnetic ordering transition. 

 We model the spin Hall insulator by the Kane-Mele Hamiltonian  with U(1) spin symmetry \cite{KM-05}.  The helical edge state is protected  by time  reversal symmetry that prohibits single particle backward scattering.  Furthermore, the U(1) spin symmetry forbids two(and higher)-particle backward scattering. As a consequence,  even in the presence of  strong correlations that do not close the bulk gap, the edge is characterized by a ``helical" Luttinger liquid (LL).\cite{HA-13}

\begin{figure}
  \centering
  \includegraphics[width=\linewidth]{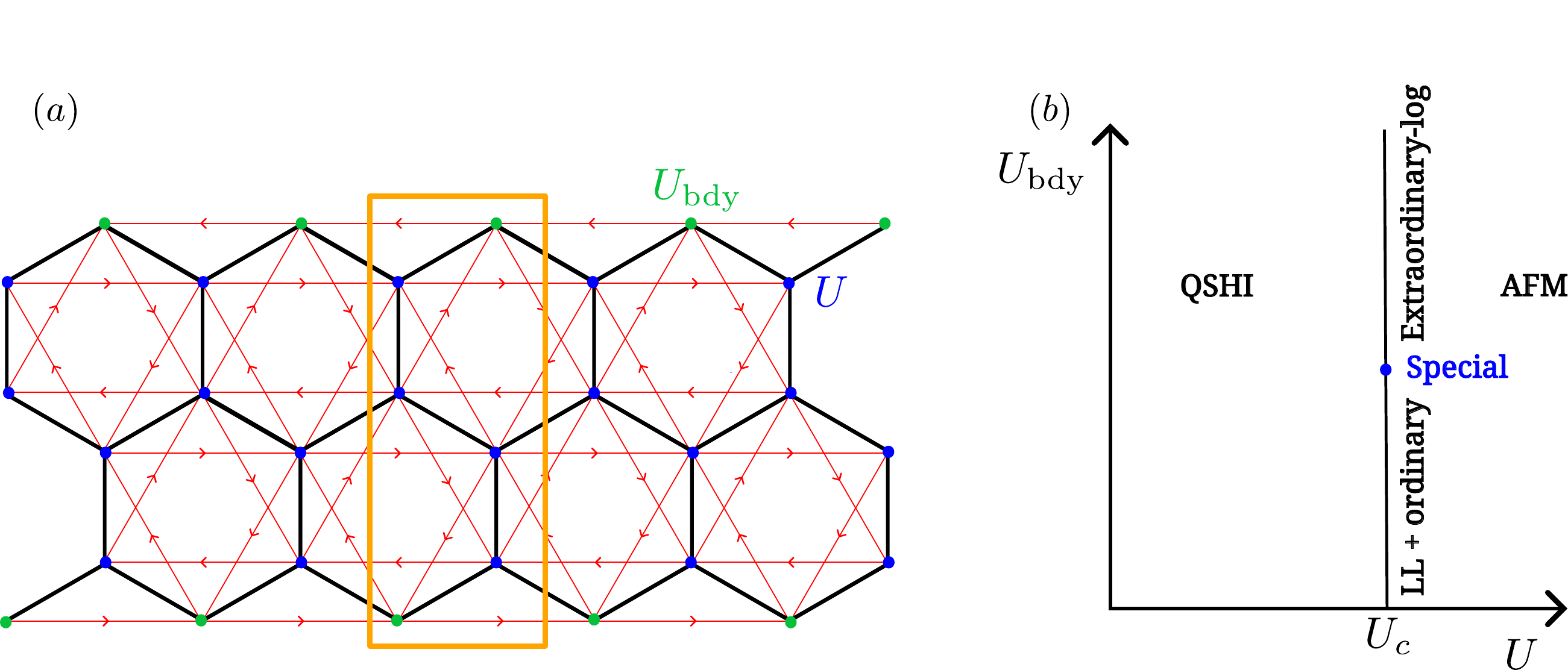}
  \caption{(a) Illustration of the geometry for $L=5$ and $L_\perp=3$.
  The nearest-neighbor hopping terms $t$ are indicated with a black line. Oriented red lines indicate the next-nearest neighbor hopping terms $i\lambda$ for the spin $\uparrow$ sector; the corresponding hoppings  for the spin $\downarrow$ sector have inverted arrows.
  Different colors distinguish the onsite Hubbard repulsion $U_{\text{bdy}}$ on the edges and $U$ in the bulk.
  A rectangle encloses the elementary unit cell. (b) Boundary/Bulk phase diagram of the model.
  For $U<U_c$ the model realizes a quantum spin Hall insulating phase, while for $U>U_c$ it is an XY antiferromagnet.
  On the bulk critical line $U=U_c$, as a function of $U_{\text{bdy}}$, we distinguish the ordinary and extraordinary-log boundary phases, and the special transition between them.
  }
  \label{fig:lattice_phasediagram}
\end{figure} 
When bulk correlations in the form of a Hubbard-U term are included, it is known that the model undergoes a three-dimensional (3D) XY transition to  a magnetically ordered phase \cite{HLA-11,PTHAH-14}.  The  ordered phase has broken time reversal symmetry and  the  edge state acquires a gap. The aim of this paper is to understand the fate of the edge state when the interaction strength  is tuned to the bulk critical point.  We  will show that as a function of the strength of the  boundary Hubbard-U interaction, a boundary quantum phase transition occurs between an ordinary phase where the helical edge  decouples from the bulk \cite{GV-12}  and an extraordinary-log phase. The extraordinary-log boundary phase was originally  predicted to be realized on the boundary of 3D classical and 2+1D quantum systems undergoing a bulk phase transition in the $O(N)$ universality class (UC), and is characterized by order parameter correlations that decay slowly as a power of logarithm of separation and an order parameter stiffness that grows logarithmically in system size.\cite{Metlitski-20, PKMGM-21, KM-23} The existence of this phase has been numerically confirmed in classical and quantum bosonic models in Refs.~\cite{PT-20,HDL-21,PTM-21,SLL-22,SHL-23,PTKM-24}, further, it is now understood that the extraordinary-log boundary order is quite ubiquitous and is the closest that the two-dimensional boundary of a 3D CFT can generically come to spontaneously breaking a continuous symmetry.\cite{CZ-24} Our finding of a boundary phase transition in the Kane-Mele-Hubbard model confirms previous theoretical predictions made in related contexts \cite{Metlitski-20, ZhangExciton, MWX-24} and is the first time that the extraordinary-log  phase is numerically observed in a system of fermions. Importantly, the two boundary phases that we find have  distinct signatures in the single particle  spectral function  thereby  allowing  detection using scanning tunnelling microscopy on the edge.  A sketch summarizing our  results is shown in Fig.~\ref{fig:lattice_phasediagram}(b).

\prlsection{Model}
We study the Kane-Mele-Hubbard model on
the honeycomb lattice, imposing periodic boundary conditions (BCs) in one direction and open BCs in the other, so as to realize two zigzag edges.
The Hamiltonian reads
\begin{equation}
\begin{split}
  {\cal H} = 
  &-t\sum_{\<\vec{\imath},\vec{\jmath}\>,\sigma} \cdag{\vec{\imath},\sigma}\cd{\vec{\jmath},\sigma}
  +i\lambda\sum_{\<\<\vec{\imath},\vec{\jmath}\>\>,\sigma} \cdag{\vec{\imath}}(\vec{\nu}^{\phantom{\dagger}}_{\vec{\imath},\vec{\jmath}} \cdot \vec{\sigma})\cd{\vec{\jmath}} \\
 &+ U_{\text{bdy}}\sum_{\vec{\imath}\in \partial}\left(\nop{\vec{\imath},\uparrow}-\frac{1}{2}\right)\left(\nop{\vec{\imath},\downarrow}-\frac{1}{2}\right) \\
 &+ U\sum_{\vec{\imath}\notin \partial}\left(\nop{\vec{\imath},\uparrow}-\frac{1}{2}\right)\left(\nop{\vec{\imath},\downarrow}-\frac{1}{2}\right),
\label{kmh_zigzag}
\end{split}
\end{equation}
where $\cdag{\vec{\imath},\sigma}$ and $\cd{\vec{\imath},\sigma}$ are the creation and annihilation operators of an electron at lattice site $\vec{\imath}$ with spin $\sigma$, and $\nop{\vec{\imath},\sigma} \equiv \cdag{\vec{\imath},\sigma}\cd{\vec{\imath},\sigma}$ is the corresponding number operator.
In Eq. (\ref{kmh_zigzag}), the first term describes single-particle hopping between nearest neighbor sites on the honeycomb lattice, the second term represents the spin-orbit interaction, leading to hoppings between next-nearest neighbor sites.
For a hopping process between the site $\vec{\imath}$ and $\vec{\jmath}$, $\vec{\nu}^{\phantom{\dagger}}_{\vec{\imath},\vec{\jmath}} = (\vec{r}-\vec{\imath})\times (\vec{\jmath}-\vec{r}) / |(\vec{r}-\vec{\imath})\times (\vec{\jmath}-\vec{r})| = \pm \vec{e}_z$, where $\vec{r}$ is the lattice site which is nearest neighbor to $\vec{\imath}$ and $\vec{\jmath}$, and $\vec{e}_z$ the unit vector in the $z-$direction.
Without loss of generality, here and in the following we fix the units by setting $t=1$.
The last two terms in Eq.~(\ref{kmh_zigzag}) represent an onsite Hubbard repulsion.
Here, we distinguish Hubbard interaction constants $U_{\text{bdy}}$ on the edges, and $U$ in the inner bulk 

In Fig.~\ref{fig:lattice_phasediagram}(a) we illustrate the geometry of the lattice.
We indicate with $L$ the number of elementary unit cells, equivalent to the number of the ``tips'' on one edge.
The elementary unit cell has $N_{\text{orb}} = 2 L_\perp$ sites, where
$L_\perp$ indicates the number of oblique edges along the unit cell.
In this way we realize the zigzag edge.
In the simulations we have $L$ even and $L_\perp = L+1$: this condition avoids a spurious finite-size gap on the edge modes \cite{CLC-14}.
The total size of the lattice is $LN_{\text{orb}} = L(2L+2)$.

In the thermodynamic limit, the boundary
 corrections are subleading and the quantum critical behavior is controlled by  $\lambda$ and $U$.
For $\lambda=0$ one realizes the Hubbard model on the honeycomb lattice, which exhibits a quantum critical point in the chiral Heisenberg Gross-Neveu UC \cite{Herbut-06,HJV-09,HJR-09,PTHAH-14}.
The inclusion of a spin-orbit term opens a mass gap at the Dirac points, resulting in a correlated quantum spin Hall insulator \cite{KM-05,KM-05b,HA-13,PTHAH-14}, which is stable for small values of $U$.
On increasing $U$ the ground state exhibits a quantum phase transition to a Neel antiferromagnet with spins pointing in the $xy$ plane, whose critical behavior belongs to the classical 3D XY UC \cite{PTHAH-14}.
We have simulated the model with finite temperature auxiliary-field QMC method \cite{BSS-81,AF_notes}, using the ALF package \cite{ALF,ALF_v2}. Spectral functions are computed using the ALF implementation of the stochastic analytical continuation method \cite{Beach-04,Sandvik-98}.
Using standard finite-size scaling (FSS) techniques, in Appendix A we determine the location of the bulk quantum critical point for $t=1$, $\lambda=0.2$ to be $U=U_c=5.723(1)$.

\begin{figure}
    \centering
    \includegraphics[width=0.9\linewidth]{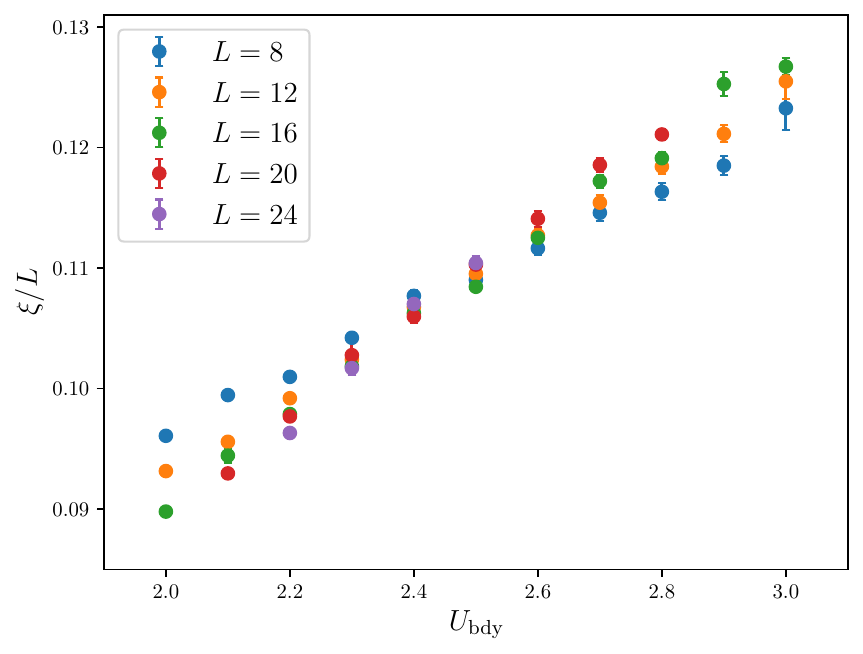}
    \caption{RG-invariant $\xi/L$ on the boundary, for a critical bulk and as a function of the edge coupling constant $U_{\text{bdy}}$.}
    \label{fig:xil}
\end{figure}
To realize the various edge phases,
we simulate the model fixing $\lambda=0.2$, $U=5.723$ at the bulk quantum critical point, and tuning the boundary parameter $U_{\text{bdy}}$.
In all simulations we have fixed the inverse temperature $\beta$ equal to the lattice size $L$, consistent with a dynamical exponent $z=1$. 
To investigate the onset of a boundary phase transition, we have computed the Renormalization-Group (RG) invariant ratio $\xi/L$, where $\xi$ is the finite-size correlations length on the boundary, defined as \cite{CP-98,PTHAH-14}
\begin{equation}
\left(\frac{\xi}{L}\right)^2 = \frac{1}{(2 L \sin(\pi /L))^2} \left(\frac{\tilde{C}(0)}{\tilde{C}(2\pi/L)} - 1\right),
\label{xildef}
\end{equation}
with $\tilde{C}(\vec{p}) = \sum_x C(x) e^{-i px}$ the Fourier transform of the boundary transverse spin correlations $C(x) \equiv \sum_{\alpha=x,y}\< S^{(\alpha)}_0 S^{(\alpha)}_x\>$.
In Fig.~\ref{fig:xil} we show $\xi/L$ at the bulk critical point, and as function of $U_{\text{bdy}}$ for lattice sizes $8\le L\le 24$.
We observe a crossing at $U_{\text{bdy}}=U_{\text{bdy},c} \approx 2.5$, indicative of a boundary phase transition between an ordinary phase at $U_{\text{bdy}}<U_{\text{bdy},c}$ and an extraordinary phase at $U_{\text{bdy}}>U_{\text{bdy},c}$.
 We review the theory of the two edge phases and the transition in more detail in the Supplemental Material (SM). Briefly, in the quantum spin Hall phase, the zigzag edge hosts a helical LL  where spin correlations decay as $\<S^{(x)}_0S^{(x)}_x\> \sim 1/x^{2K}$, with $K$ the LL parameter.
On tuning the bulk to the critical point, the most relevant boundary interaction can be schematically written as
\begin{equation}
    S_{\text{int}} \sim \int dx d\tau \vec{S}(x,\tau)\vec{\hat{\phi}}(x,\tau)
\label{bulk_bdy_int}
\end{equation}
where $\vec{\hat{\phi}}$ is the boundary field operator of the XY UC with ordinary BCs, with scaling dimension $\Delta_{\hat{\phi}} = 1.2286(25)$\cite{PT-23}.
For a sufficiently small value of $U_{\text{bdy}}$, $K > 2 - \Delta_{\hat{\phi}} \approx 0.77$ and the interaction is irrelevant, leading to a LL decoupled from the ordinary boundary. Such an edge mode decoupling has been noted in a number of topological systems undergoing a bulk symmetry breaking transition in Ref.~\cite{GV-12}.

\begin{figure}[t]
  \includegraphics[width=\linewidth]{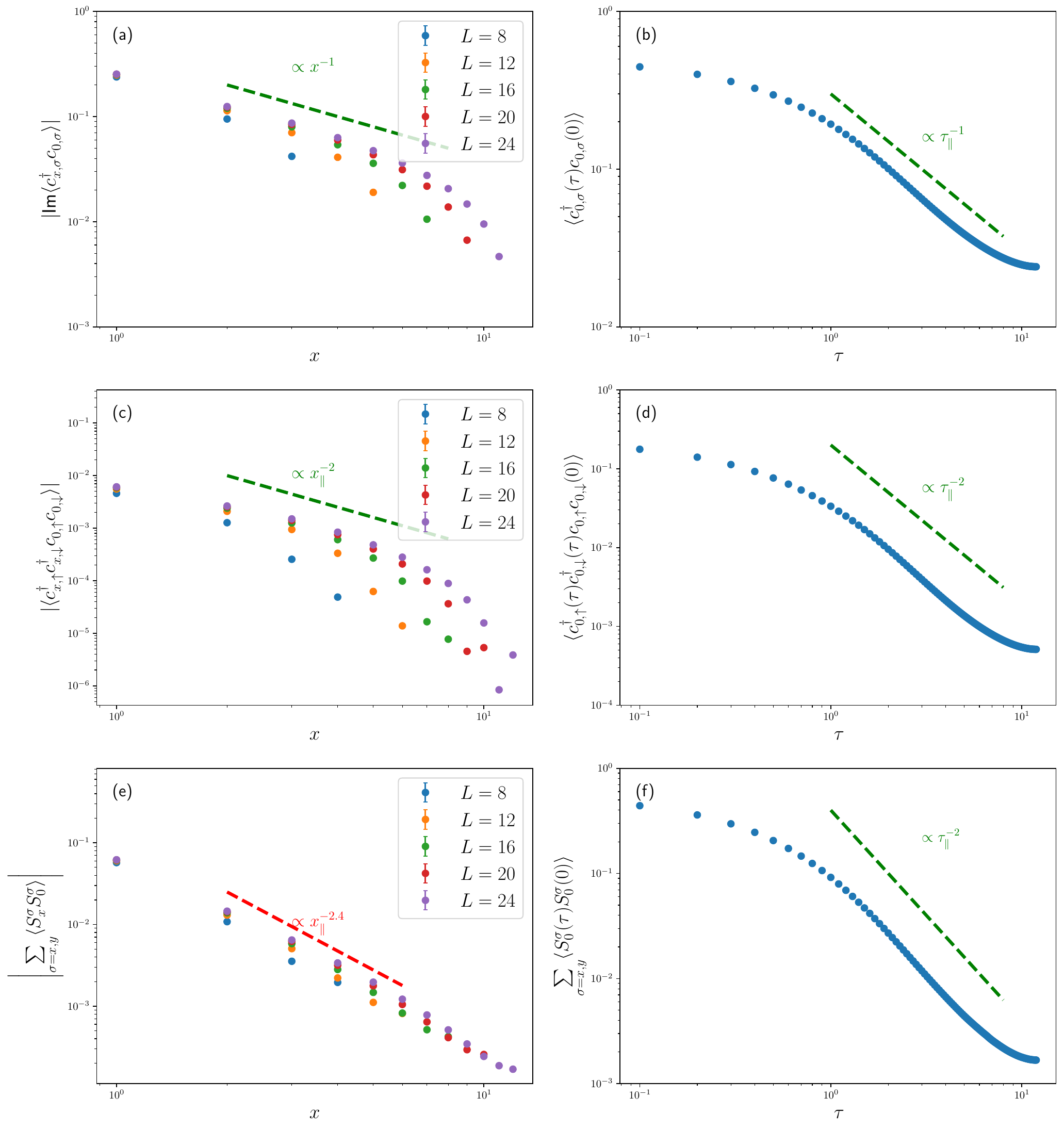}
   \caption{ Edge single particle, Cooper pair and transverse spin correlation functions along the space and time directions in the ordinary phase at
   $U_{\rm bdy}=0$. }
   \label{fig:green_pair}
 \end{figure}

We have studied the ordinary phase by setting $U_{\text{bdy}} = 0$.
Here the edge state corresponds to a helical LL and we can extract the LL parameter $K$ by analyzing the space/imaginary time  decay of the single particle and Cooper pair $\Delta_i = c_{i \uparrow} c_{i \downarrow}$ correlation function\cite{SM}.  As shown in Fig. \ref{fig:green_pair} 
the data are consistent  with  $\text{Im} \langle \hat{c}^{\dagger}_{x}(\tau)\hat{c}^{\phantom\dagger}_{0}(0) \rangle \propto \frac{1}{\sqrt{(v_F \tau)^2 + x^{2} }}  $ and  $\langle \hat{\Delta}^{\dagger}_{x}(\tau)\hat{\Delta}^{\phantom\dagger}_{0}(0) \rangle \propto \frac{1}{(v_F \tau)^2 + x^{2} } $,  consistent with
Luttinger parameter $K \approx 1$  as for the non-interacting Kane-Mele model.
For the transverse spin-spin correlations we expect contributions both from the edge LL mode with scaling dimension $\Delta_{\vec{S}} = K$ and from the boundary order parameter of the ordinary UC $\vec{\hat{\phi}}$ with scaling dimension $\Delta_{\hat{\phi}}$. However, for $K \approx 1$ it is difficult to distinguish these contributions for our lattice sizes. Further, for $K \approx 1$ the interaction in Eq.~(\ref{bulk_bdy_int}) is only slightly irrelevant, thus, our results in the ordinary phase may be affected by corrections to scaling.

\begin{figure}
    \centering
    \includegraphics[width=\linewidth]{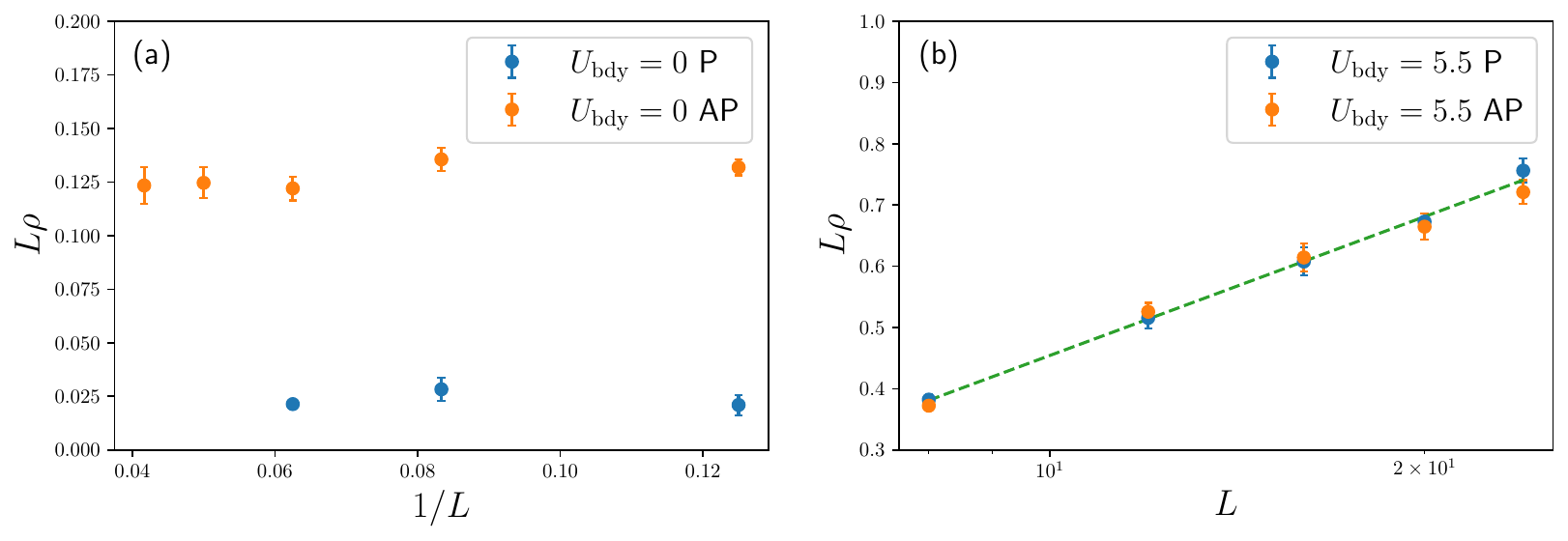}
    \caption{Spin stiffness at the ordinary (a) and extraordinary-log (b) phases, for periodic (P) and antiperiodic (AP) lateral BCs.
    A dashed line indicate a linear fit of $L\rho$ to $\ln L$.
    }
    \label{fig:stiffnesss}
\end{figure}

To analyze the extraordinary phase, we have simulated the model at $U_{\text{bdy}} = 5.5$, and computed the spin stiffness $\rho$, which is obtained by considering a twist of an angle $\varphi$ in the lateral direction.
Then the stiffness $\rho$ is defined as the response of the system to the torsion \cite{FBJ-73,SM,rhonorm}
\begin{equation}
    \rho = - \frac{1}{\beta} \frac{d^2}{d\varphi^2}\ln Z\Big|_{\varphi = 0}.
    \label{rhodef}
\end{equation}
With this definition $L\rho$ is a RG invariant observable.
A comparison of $L\rho$ in the ordinary and extraordinary phase shown in Fig.~\ref{fig:stiffnesss} reveals a striking difference: while in the ordinary phase $L\rho$ appears to converge to a constant for $L\rightarrow \infty$, in the extraordinary phase we observe a logarithmic growth of $L\rho$.
Such logarithmic violation of standard FSS is a peculiar feature of the extraordinary-log phase \cite{Metlitski-20,PTKM-24}.

\begin{figure}
    \centering
    \includegraphics[width=0.9\linewidth]{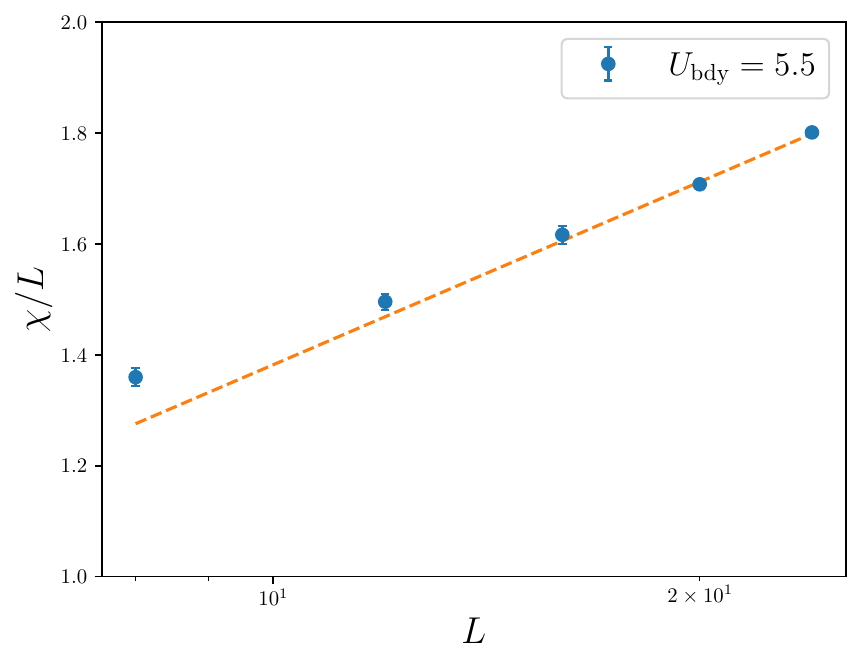}
    \caption{Spin susceptibility in the  extraordinary-log phase for periodic lateral BCs.
    A dashed line indicates a linear fit of $\chi/L$ to $\ln L$.
    }
    \label{fig:susc}
\end{figure}

For a quantitative analysis, we recall that in the extraordinary-log phase the boundary can be to a leading approximation described by a LL with a Luttinger parameter $K$, which flows logarithmically to the ${\rm XY}$ ordered fixed point $K=0$ \cite{Metlitski-20}. Matching to the notation of Ref.~\cite{Metlitski-20}, we write $K$ in terms of the coupling constant $g$, $K  = \frac{g}{4\pi}$. The  $\beta$-function of $g$ for $g \to 0$ is given by
\begin{equation} \beta(g) = -\frac{dg}{d \ell} \approx  \alpha_r g^2, \quad\quad \alpha_r = \alpha \frac{1}{2}\left(v_s/v_b + v_b/v_s\right), \label{alphar} \end{equation}
where $v_s$ and $v_b$ are the surface and bulk velocities, and $\alpha$ is a universal parameter of the extraordinary-log phase estimated to be $\alpha =0.300(5)$ \cite{PTM-21}.
The velocity ratio $v_s/v_b$ flows under the RG slowly to $1$, as a power of $\ln L$ \cite{Metlitski-20}, such that
for the feasible range of lattice sizes $v_s/v_b$ is effectively constant, resulting in a correction factor to $\alpha$ in Eq.~(\ref{alphar}).  To extract the surface velocity, in addition to $\rho$ we compute the total spin-susceptibility of the system,  $\chi \equiv \beta \< (\sum_i S_i^{(z)})^2\>$, see Fig.~\ref{fig:susc}. In a standard single channel LL with $g \to 0$, $\rho L \approx \frac{v_s}{g}$, $\chi/L \approx \frac{1}{g v_s}$ (see Appendix B). Replacing $g \to g(L)$ to account for the flow of $g$ and including the contribution from both edges of the system,
we fit $L\rho$ and $\chi/L$ to
\begin{align}
    \chi / L &= A + B \ln L,
    \label{fit_chi}\\
    L\rho &= C + D \ln L,
    \label{fit_rho}
\end{align}
where
\begin{equation} v_s = \sqrt{D/B}, \quad\quad  \alpha_r = \sqrt{DB}/2. \label{valpharDB}\end{equation}
Fits of Eqs.~(\ref{fit_chi}) and (\ref{fit_rho}) allow us to estimate $v_s \approx 0.9(1)$ and $\alpha_r \approx 0.2$.
The estimate of $v_s$ is also confirmed by fitting the expected exponential decay of the imaginary time correlations on the edge $\sum_{\sigma = x,y}\langle \tilde{S}^\sigma_{k_{\text{min}}}(\tau) \tilde{S}^\sigma_{-k_{\text{min}}}(0) \rangle$, at the minimum nonzero momentum \cite{SM}.
By fitting $\sum_{\sigma = x,y}\langle \tilde{S}^\sigma_{k_{\text{min}}}(\tau) \tilde{S}^\sigma_{-k_{\text{min}}}(0) \rangle$ to $A \exp\{-k_{\rm min} v_s \tau\}$, we extract $v_s=0.67(20)$, in agreement within precision with the previous estimate.
Analogously, we estimate the bulk velocity $v_b$ from imaginary-time correlations, obtaining $v_b \approx 0.68(5)$.
Using these estimates of $v_s$ and $v_b$, the
renormalization factor in Eq.(\ref{alphar}) is $(v_s/v_b + v_b/v_s)/2 = 1.04(10)$, i.e., $1$ within precision.
  Considering $\alpha_r\simeq \alpha$, the value of $\alpha$ extracted from Eq.~(\ref{valpharDB})
  differs considerably from the expected value $\alpha=0.300(5)$ \cite{PTM-21}.
This can be explained with the presence of subleading corrections to the RG flow. The  $\beta$-function of $g$ can be expanded beyond the leading order as
  \begin{equation}
    \beta(g) = \alpha_r g^2 + b g^3 + O(g^4),
    \label{beta}
  \end{equation}
where the universal parameter $b$ has been estimated as $b=-0.03(1)$ \cite{PTKM-24} for $v_s/v_b = 1$ --- we use this velocity ratio in the estimates below.  Thus, $\frac{d}{d\ell} \left(\frac{1}{g}\right) \approx \alpha + b g$, suggesting that
  for a finite not-so-small value of $g$, over a finite range of lattice sizes one observes an effective $\alpha$ which is given by $\alpha_{\rm eff} \approx \alpha + b g$.
  We can estimate the value of $g$ in the range of available data by inspecting Fig.~\ref{fig:stiffnesss} (b), where we
  observe $0.4\lesssim L\rho\lesssim 0.8$.
  This is related to $g$ by $L\rho \approx 2 v_s/g$.
  Having estimated $v_s=0.9(1)$, in the range of available QMC data we have $2 \lesssim g \lesssim 4$.
  In this interval $\alpha_{\rm eff}$ is roughly $0.3-3 \cdot 0.03(1) \approx 0.2$, on spot with the value of $\alpha$ as extracted from Eq.~(\ref{valpharDB}).

A curious difference between the ordinary and extraordinary-log phases illustrated in Fig.~\ref{fig:stiffnesss} is the sensitivity to fermion BCs along the latteral direction. In the ordinary phase at $U_{\rm bdy} = 0$ we observe a significant difference in the stiffness $\rho L$ for periodic and anti-periodic fermion BCs, while in the extraordinary-log regime the difference is negligible. This is in qualitative  agreement with the  expectation. First, the entire dependence on the fermion  BCs comes from the edge, as the bulk is insulating.  In the ordinary phase, the decoupled edge LL  responds to a change in the fermion BCs. An explicit calculation (see Appendix B)  predicts at $U_{\rm bdy} = 0$, $(\rho(A) - \rho(P)) L \approx 0.08$, compared to  $(\rho(A) - \rho(P)) L \approx 0.100(6)$ observed at the largest system size $L =16$, where we have MC data for both periodic and antiperiodic BCs.  Here in the theoretical estimate we've used $K = 1$ and the  edge velocity $v_s \approx 0.8$, extracted for $U_{\rm bdy} = 0$ from the imaginary time decay of the electron Green's function along the edge at the smallest momentum $k_{\text{min}} = \pi + \frac{2\pi}{L}$ (see SM). In the extraordinary-log phase, we expect a negligible dependence on the fermion BCs just as is the case for a LL with $K \to 0$ and $v_s \beta/L$ -- fixed, see Appendix B.

\begin{figure}[t]
  \includegraphics[width=\linewidth]{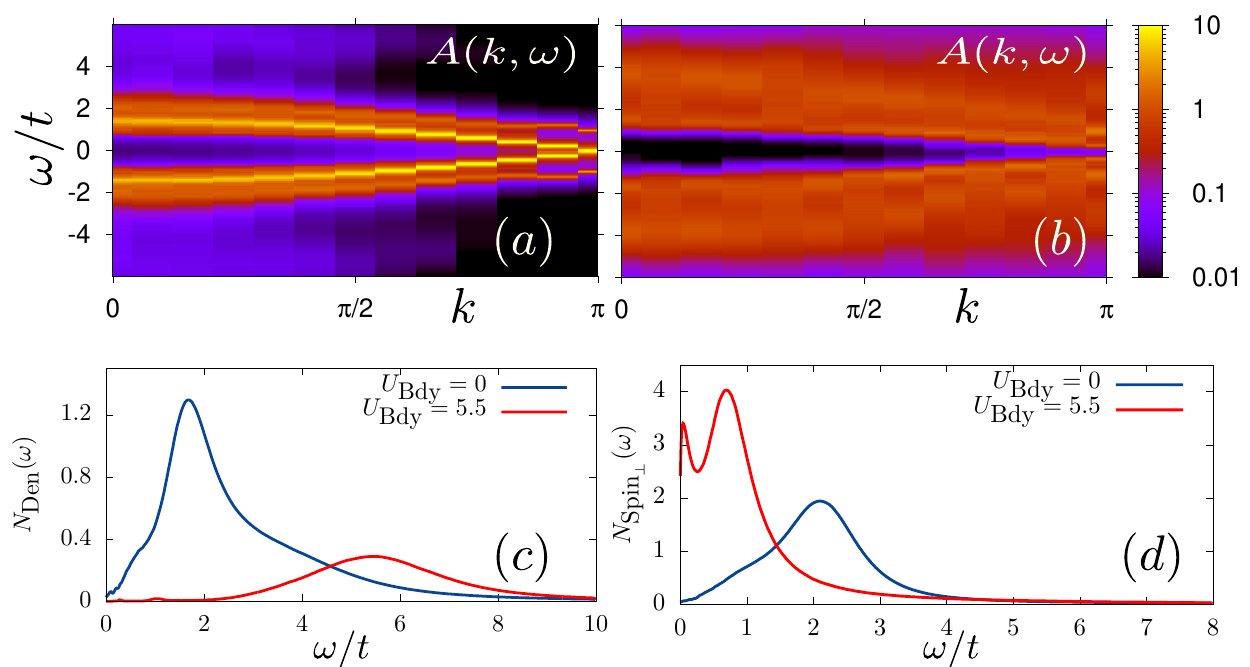}
   \caption{ Single particle edge spectral function  on an $L=24$ lattice at $\beta t = 24$ at criticality, $U_c/t = 5.723$  for  (a) $U_{\rm bdy} = 0 $  and (b) $U_{\rm bdy} = 5.5 $. Local  spectral  function for (c) density fluctuations  and for (d) transverse spin fluctuations.   Here, we again consider $L=24$ and $\beta t = 24$. }
   \label{fig:spectral}
 \end{figure}

\prlsection{Spectral signatures} The ordinary and extraordinary-log phases have distinct spectral signatures in the particle  and particle-hole channels.  Consider the single particle  edge spectral  function, 
$ A(k,\omega) = -\frac{1}{\pi} \text{Im} 
\int_{0}^{\infty}  d t  e^{i \omega t} \left< \left\{ \hat{c}^{\phantom\dagger}_{k}(t),  \hat{c}^{\dagger}_{k}(0)\right\} \right> $ with  $ \hat{c}^{\dagger}_{k} =  \frac{1}{\sqrt{L}} \sum_{\vec{\imath} \in \partial } e^{i k \vec{e}_x \cdot \vec{\imath}} \hat{c}^{\dagger}_{\vec{\imath}} $. 
   In the ordinary phase,  $U_{\rm bdy} = 0$   the edge state corresponds to a helical LL with as discussed above, a Luttinger  parameter $K$ close to unity akin to a non-interacting helical edge mode. 
This leads to the  well defined edge mode in the spectral function seen in Fig.~\ref{fig:spectral}(a). In contrast, in the extraordinary-log phase the spectral function in Fig.~\ref{fig:spectral}(b) displays a continuum spectrum, similar to that of a LL with small $K$. Qualitatively, in the  extraordinary-log phase, the transverse spin-spin correlations  decay  very slowly  and on our finite system sizes  mimic long-range order.  This leads to spin-flip scattering between  the left spin-up and right spin-down electrons of the helical edge thus opening a pseudo-gap\cite{MWX-24}.  The local spectral function in the  particle-hole channel, $N_{O} (\omega) = -\frac{1}{\pi} \text{Im}  \frac{1}{L} \sum_{\vec{\imath} \in \partial}
\int_{0}^{\infty}  d t  e^{i \omega t} \left< \left[ \hat{O}_{\vec{\imath}}(t),  \hat{O}_{\vec{\imath}}(0)\right] \right> $,  is depicted in Fig.~\ref{fig:spectral}(c) for the charge $\hat{O}_{\vec{\imath}} = \sum_{\sigma} \hat{n}_{\vec{\imath},\sigma} $   and in Fig.~\ref{fig:spectral}(d) for the spin $\hat{O}_{\vec{\imath}} =  \hat{S}_{\vec{\imath}}^{x} $ channels.   In the  extraordinary-log phase ($U_{\rm bdy}=5.5$), in  contrast to the ordinary case ($U_{\rm bdy} = 0$), the quasi long range order  depletes low lying charge modes while  spin fluctuations acquire substantial low lying weight.  These pseudo-gap features in the  single particle and charge sectors provide distinct spectral signatures of the 
extraordinary-log phase that can be picked up with local experimental probes. 

\prlsection{Conclusions}
The Kane-Mele-Hubbard model  provides a unique possibility to investigate numerically  boundary criticality in topological phases of matter.  The bulk transition belongs to the 3D XY UC.
Upon tuning the value of the boundary Hubbard-U term  we observe  both an ordinary edge phase coexisting with a helical LL decoupled from the bulk and an extraordinary-log phase  with logarithmically diverging spin stiffness.  Our results provide the first realization of the extraordinary-log  phase in  a fermionic model. A detailed analysis of the result  provides a connection with  previous analytical and numerical studies of classical 3D XY  boundary criticality.  The key differences between our model and the classical boundary criticality include: i) the presence of the decoupled LL mode  in the ordinary phase, ii) the nature of the transition between the ordinary and the extraordinary-log phases: unlike in the classical case, as reviewed in the SM this transition is under complete theoretical control\cite{Metlitski-20} and features an exponentially diverging correlation length (akin to  the  Kosterlitz-Thouless transition). The reason for these differences is the absence of  XY order parameter phase slip events on the edge in the Kane-Mele-Hubbard model, which are prohibited by fermion number conservation --- a consequence of the non-trivial topology of the quantum spin Hall insulator that survives at the bulk XY critical point.

\prlsection{Note added} Upon completing this work, we became aware of Ref.~\cite{ge25} that studies a very similar setup.  The crucial difference, however,  lies in symmetry. Ref.~\cite{ge25} includes spin-orbit coupling such that the symmetry of the bulk transition is reduced from 3D XY  to 3D Ising.  Consequently, the symmetry of the helical edge state is  reduced from U(1) to  $\mathbb{Z}_2$.  The nature of the boundary phases should hence  differ. In particular, the extraordinary-log phase is not realized with Ising symmetry, instead, the large $U_{\rm bdy}$ edge phase has true long range order and fully gapped fermions.

\begin{acknowledgments}
  F.P.T. is funded by the Deutsche Forschungsgemeinschaft (DFG, German Research Foundation), Project No. 414456783.  F.F.A. thanks the DFG for financial support under the AS 120/19-1 grant (Project number, 530989922) as well as the W\"urzburg-Dresden
  Cluster of Excellence on Complexity and Topology in Quantum Matter ct.qmat
  (EXC 2147, project-id 390858490).
  The authors gratefully acknowledge the Gauss Centre for Supercomputing e.V. \cite{gauss} for funding this project by providing computing time on the GCS Supercomputer SuperMUC-NG at Leibniz Supercomputing Centre \cite{lrz}.

\end{acknowledgments}

\bibliography{francesco,extra}

\begin{thebibliography}{68}
\bibitem [{rho()}]{rhonorm}%
  \BibitemOpen
  \href@noop {} {}\bibinfo {note} {Note, our definition of the stiffnes here
  differs from the standard definition $\rho = - \frac{1}{\beta}
  \frac{L_x}{L_y}\frac{d^2}{d\varphi^2}\ln Z\Big|_{\varphi = 0}$ by the aspect
  ratio of the strip $L_y/L_x$.}\BibitemShut {Stop}%
\end{thebibliography}

\onecolumngrid
  \parbox[c][4em][c]{\textwidth}{\centering \large\bf End Matter}
\smallskip
\twocolumngrid

\prlappendix{Bulk critical point}
\begin{figure}
    \centering
    \includegraphics[width=0.9\linewidth]{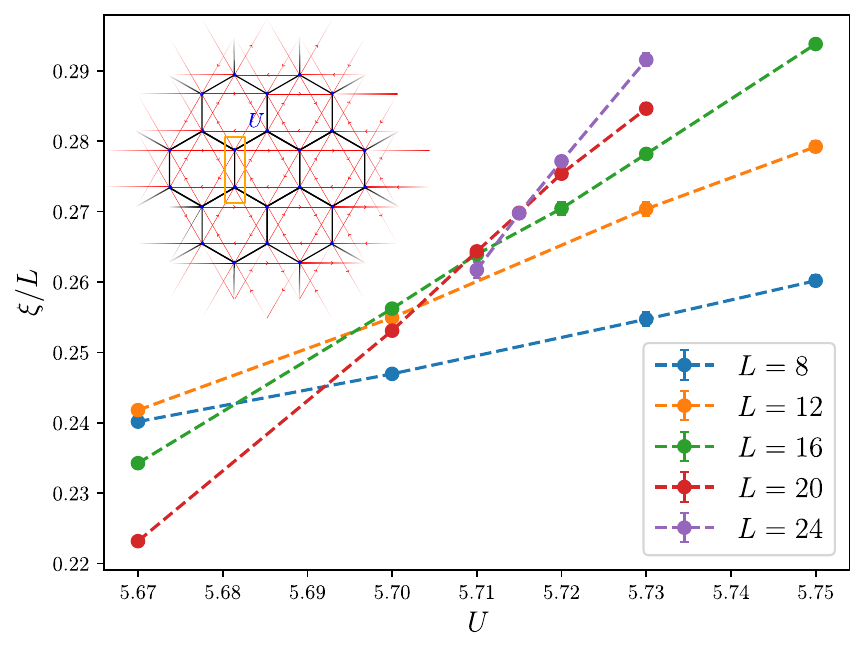}
    \caption{RG-invariant ratio $\xi/L$ for periodic BCs, close to the quantum critical point. Dashed lines are a guide to the eye.
      The inset illustrates the lattice. Color and arrow conventions are as in Fig.~\ref{fig:lattice_phasediagram}.
      A rectangle encloses the elementary unit cell.
    }
    \label{fig:xil_bulk}
\end{figure}
To determine the bulk critical point, we study the Kane-Mele-Hubbard model
on a honeycomb lattice with $L\times L$ unit cells, applying periodic BCs on both directions.
The lattice is illustrated in the inset of Fig.~\ref{fig:xil_bulk}.
The Hamiltonian is
\begin{multline}
  {\cal H} = 
  -t\sum_{\<\vec{\imath},\vec{\jmath}\>,\sigma} \cdag{\vec{\imath},\sigma}\cd{\vec{\jmath},\sigma}
  +i\lambda\sum_{\<\<\vec{\imath},\vec{\jmath}\>\>,\epsilon\epsilon'} \cdag{\vec{\imath},\epsilon}(\vec{\nu}^{\phantom{\dagger}}_{\vec{\imath},\vec{\jmath}} \cdot \vec{\sigma})_{\epsilon\epsilon'}\cd{\vec{\jmath},\epsilon'} \\
 + U\sum_{\vec{\imath}}\left(\nop{\vec{\imath},\uparrow}-\frac{1}{2}\right)\left(\nop{\vec{\imath},\downarrow}-\frac{1}{2}\right).
\label{kmh}
\end{multline}
The location of the quantum critical point is obtained by a FSS analysis of the RG-invariant ratio $\xi/L$, where $\xi$ is defined by Eq.~(\ref{xildef}), employing the spatial correlations of the order parameter $\vec{S} = (S^{(x)},S^{(y)})$.
In Fig.~\ref{fig:xil_bulk} we plot $\xi/L$ for values of $U$ close to the quantum critical point, and available lattice sizes $8\le L \le 24$.
According to RG, in the vicinity of a critical point $U=U_c$, $\xi/L$ should scale as
\begin{equation}
    \xi / L = f((U-U_c)L^{1/\nu}) + L^{-\omega}g((U-U_c)L^{1/\nu}),
    \label{xil_scaling}
\end{equation}
where $1/\nu = 1.48864(22)$ \cite{CLLPSDSV-19} is the leading relevant exponent and we have included the leading correction to scaling.
For a quantitative determination of $U_c$, following a standard procedure \cite{PTPV-09} we Taylor-expand the right-hand side of Eq.~(\ref{xil_scaling})
\begin{equation}
    \xi / L = \sum_{n=0}^m f_n ((U-U_c)L^{1/\nu})^n + L^{-\omega}\sum_{n=0}^k g_n ((U-U_c)L^{1/\nu})^n,
\label{xil_scaling_taylor}
\end{equation}
and fit QMC data to the right-hand side of Eq.~(\ref{xil_scaling_taylor}).
Crucially, the critical exponent $\nu$ and the leading irrelevant exponent $\omega$ are known with a great accuracy $1/\nu = 1.48864(22)$ \cite{CLLPSDSV-19}, $\omega = 0.789(4)$ \cite{Hasenbusch-19}.
This allows us to input the value of $\nu$ and $\omega$ and fit the amplitudes only, enabling us to accurately estimate $U_c$.
A truncation of the expansion (\ref{xil_scaling_taylor}) to $m=1$ and $k=0$ adequately describes the QMC data.

Fits reported in the SM deliver a very stable value of $U_c$, with a good \chidof.
A variation of the critical exponents $\nu$ and $\omega$ within the available accuracy does not significantly alter the fitted amplitudes.
Based on the fit results we conservatively estimate
\begin{equation}
    U_c = 5.723(1).
\end{equation}
We observe that the leading correction-to-scaling exponent $\omega$ used in the analysis corresponds to the lowest irrelevant operator, scalar with respect to the $O(2)$ and conformal symmetries.
Indeed, this is the leading correction observed in classical lattice models \cite{PV-02}.
In the present quantum model, additional irrelevant operators that break the Lorentz symmetry are in principle present.
Their scaling dimension is however significantly larger \cite{MSS-17,Henriksson-23}, such that scaling corrections in the QMC data are reliably captured by the leading irrelevant scalar.
The fits of Table \ref{tab:fits_bulk} confirm this analysis.

\prlappendix{Spin stiffness and susceptibility}
In this appendix we discuss the edge contribution to the spin susceptibility and stiffness. 
The Lagrangian of the Kane-Mele-Hubbard model in imaginary time can be schematically written as
\begin{multline}
{\cal L}= \sum_{i}\pdag{i}\left(\partial_\tau - i A_{i}^{(0)}(\tau) \frac{\sigma^3}{2}\right) \pd{i} + \sum_{ij}   \pdag{i}  T_{ij} e^{i A _{ij} \frac{\sigma^3}{2}} \pd{j} \\
+ U \sum_i \left(\pdag{i}\pd{i} - \frac{1}{2}\right)^2,
\label{lagrangian}   
\end{multline}
where $\{\pdag{i}, \pd{i}\}$ are the fermionic variables and for economy of notation we have suppressed the spin indices.
The hopping matrix $T$  additionally depends on the spin, allowing to describe the spin-orbit interaction.
In Eq.~(\ref{lagrangian}) we have introduced for later convenience an external gauge field $A_{\mu}$ for the $U(1)$ spin symmetry:   $A_i^{(0)}(\tau)$ is the corresponding lattice scalar potential and $A_{ij}$ is the lattice vector potential.
The Lagrangian is invariant under the gauge symmetry $\pd{i} \rightarrow e^{i \alpha_i(\tau) \sigma^3/2} \pd{i}$, $A_i^{(0)}(\tau) \rightarrow A_i^{(0)}(\tau) + \partial_\tau \alpha_i(\tau)$, $A_{ij} \rightarrow A_{ij} + \alpha_i - \alpha_j$.
Specializing Eq.~(\ref{lagrangian}) for a space- and time-independent potential $A_i^{(0)}(\tau) = A^{(0)}$,
the full susceptibility $\chi$ of the system can be computed as
\begin{equation}
\begin{split}
\chi = \beta \langle (S^z)^2 \rangle &= \sum_{ij} \int_0^{\beta} d\tau \< \pdag{i} \frac{\sigma^3}{2}  \pd{i}(\tau) \pdag{j} \frac{\sigma^3}{2} \pd{j}(0)\> \\
&= -\frac{1}{\beta} \left.\frac{d^2}{d(A^{(0)})^2} \ln Z\right|_{A=0} .
\label{chi}
\end{split}
\end{equation}  
Similarly, to compute the spin stiffness $\rho$ we introduce  twisted BCs in the $x$ direction $\pd{x+L,y} = e^{i \varphi \sigma^3/2} \pd{x,y}$, which is equivalent to choosing some vertical cut and twisting $T_{ij} \rightarrow e^{i \varphi \sigma^3/2} T_{ij}$ for every bond $(i,j)$ crossing the cut. Then $\rho$ is given by Eq.~(\ref{rhodef}).

We now proceed to the effective theory of the strip at the critical point. We begin in the ordinary edge phase, where the LL is decoupled from the bulk.  We focus on a single edge of the strip and let the boundary XY order parameter $e^{i \theta} \sim \psi^{\dagger}_{i \downarrow} \psi_{i \uparrow} = S^-_i$. The LL action takes the form
\begin{equation}
S = \frac{1}{2g}  \int dx d \tau  \Big[\frac{1}{v_s}(\partial_\tau \theta - A^{(0)})^2 
+ v_s (\partial_x \theta - A^{(1)})^2\Big].
\label{effaction}
\end{equation}
The coupling to $A^{(0)}$, $A^{(1)}$ is set by the quantum numbers of $e^{i\theta}$ ($S^z = -1$). $j^{c}_{\mu} = \frac{1}{2\pi} \epsilon_{\mu \nu} \partial_{\nu} \theta$ is the electric charge current. Turning on a flat $A^{(1)} = \frac{\varphi}{L}$ and passing to the  Hamiltonian,
\begin{equation} \frac{H}{v_s} = \frac{2\pi}{L} \left[K (S^z)^2 + \frac{1 }{4K }(N -\frac{\varphi}{2\pi})^2\right] + \sum_{k\neq 0} |k| a^{\dagger}_k a_k. \end{equation} 
Here $K = \frac{g}{4\pi}$ is the Luttinger parameter, $S^z$ is the total spin and $N = \frac{1}{2\pi}\int dx\, \partial_x \theta$ is the total electric charge. The quantization of $N$ and $S^z$ is set by the BCs. We always have $2 S_z \in \mathbb Z$ and $N \in \mathbb{Z}$. Further, when the electrons have antiperiodic BCs (before the spin twist $\varphi$ is imposed), $2 S_z = N\,\, ({\rm mod} \,\,2)$. When the electrons have periodic BCs, $2 S_z = N+1 \,\, ({\rm mod} \,\,2)$. Note that changing $\varphi \to \varphi + 2\pi$ effectively toggles between these BCs, as it should. 

The contribution of the edge to the susceptibility and stiffness is now easy to calculate. The partition function
\begin{equation} Z(\mu_s, \varphi) = {\rm tr} e^{-\beta H + \beta\mu_s S^z}, \end{equation}
where $\mu_s$ is the spin chemical potential. Evaluating the sums over $N$ and $S_z$,
\begin{eqnarray}
Z_A(\mu_s, \varphi) &=& {\cal N} e^{-|\tau|  \varphi^2/(8\pi K) }  \nonumber\\
&&\times \bigg[\nu_3\big(\frac{i |\tau| \varphi}{2 K}, e^{-2 \pi |\tau|/K}\big)\nu_3(i \beta \mu_s/2, e^{-2\pi K |\tau|}) \nonumber\\
&&+ \nu_2\bigg(\frac{i |\tau| \varphi}{2 K}, e^{-2\pi |\tau|/K}\bigg)\nu_2(i \beta \mu_s/2, e^{-2\pi K |\tau|} )\bigg], \nonumber\\
Z_P(\mu_s, \varphi) &=& {\cal N} e^{-|\tau|  \varphi^2/(8\pi K) }  \nonumber\\
&&\times \bigg[\nu_2\big(\frac{i |\tau| \varphi}{2 K}, e^{-2 \pi |\tau|/K}\big)\nu_3(i \beta \mu_s/2, e^{-2\pi K |\tau|}) \nonumber\\
&&+ \nu_3\bigg(\frac{i |\tau| \varphi}{2 K}, e^{-2\pi |\tau|/K}\bigg)\nu_2(i \beta \mu_s/2, e^{-2\pi K |\tau|} )\bigg], \nonumber\\
\label{ZAP}\end{eqnarray} 
Here $A$ and $P$ stand for periodic and antiperiodic BCs, $|\tau| = v_s \beta/L$ and ${\cal N}$ is a $\mu_s$, $\varphi$ independent constant denoting the contribution of the $k \neq 0$ modes. $\nu_i(z,q)$ are Jacobi $\theta$-functions. Then $\rho = -\frac{1}{\beta Z} \frac{d^2 Z}{d\varphi^2}\big|_{\varphi = 0, \mu_s = 0}$. For instance for $|\tau| = 0.8$, as appropriate for $\beta/L = 1$ and $v_s \approx 0.8$ extracted from the edge electron Green's function at $U_{\rm bdy} = 0$, we have $\rho(A)L/v_s \approx 0.0237$ and $\rho(P) L/v_s \approx -0.02561$. Accounting for the contribution from both edges of the strip, we obtain $(\rho(A)_{\rm strip} -\rho(P)_{\rm strip}) L \approx 0.08$ as reported in the main text. 

To compute the full stiffness of the strip in the ordinary phase (for either fermion BCs) we need to add the contributions from the 3D XY model with ordinary BCs and from  the LL. Consider the classical 3D XY model in a $T^2 \times I$ cuboid geometry, where $x$, $y$ coordinates are periodic with lengths $L_1$ and $L_2$ and the $z$ direction is open with length $L_3$. The $z = 0$ and $z = L_3$ faces carry ordinary BCs. Further, there is a twist $\varphi$ along the $x$ direction. The partition function is $Z_{{\rm ord}}(L_2/L_1, L_3/L_1, \varphi)$. Translating to our quantum model, $L_1 = L$, $L_2 = v_{b} \beta$ and $L_3 = \frac{\sqrt{3}}{2} L_1$, thus,  the 3D XY contribution to the stiffness of the quantum model is $\frac{\rho L}{v_b} = - \frac{L}{v_b \beta} \frac{d^2}{d\varphi^2}\log Z_{{\rm ord}}(\frac{v_b \beta}L, \frac{\sqrt{3}}{2}, \varphi)$. It has been estimated with MC simulations of the classical 3D XY model  that for unit aspect ratio $L_1 = L_2 = L_3$,  $-\frac{d^2}{d\varphi^2}\log Z_{{\rm ord}}(1, 1, \varphi)\approx 0.175(3)$\cite{PT-23}. This of the right order of magnitude compared to $\rho L$ in the ordinary phase in Fig.~\ref{fig:stiffnesss}, but a more quantitative analysis is precluded by the lack of knowledge of aspect ratio dependence of the stiffness of the 3D XY model.

Now, with an eye to exploring the extraordinary-log phase, let's consider the LL spin susceptibility and stiffness in the limit $K \to 0$, $|\tau|$ -- fixed. From Eqs.~(\ref{ZAP}) we find for each edge,
\begin{equation} \rho L \approx \frac{v_s}{4 \pi K}, \quad\quad \chi/L \approx \frac{1}{4 \pi v_s K} \label{Klarge},\end{equation}
with corrections that are suppressed exponentially in $1/K$. Note that the result is independent of the fermion BCs --- this is a consequence of the electric charge degrees of freedom in the LL becoming heavy as $K \to 0$. Recalling that $K$ flows logarithmically to zero in the extraordinary-log phase, we expect that the leading contribution to the stiffness and susceptibility in the extraordinary-log phase will be given by Eqs.~(\ref{Klarge}), with RG improved $K \to K(L)$.

\clearpage

%% Supplemental Material

\onecolumngrid
  \parbox[c][3em][t]{\textwidth}{\centering \large\bf Supplemental Material}
\smallskip
\twocolumngrid

\switchtoletter{S}
\section{RG description of the edge}
In this section, we review the RG treatment of the ordinary + LL and extraordinary-log phases of the Kane-Mele-Hubbard model and the special transition between them. The theory is essentially identical to that of a 2+1D Bose-Hubbard model going through a bulk insulator to superfluid transition at non-commensurate edge density, originally developed in Ref.~\cite{Metlitski-20}. This theory was applied to the quantum spin Hall - superconductor phase transition in Ref.~\cite{MWX-24}: the presently considered case of a quantum spin Hall - XY antiferromagnet transition is completely analogous. A closely related theory also describes the edge at the XY$^*$ exciton condensation transition out of a quantum Hall bilayer state in Ref.~\cite{ZhangExciton}. 

We begin in the small $U_{\rm bdy}$ regime where the ordinary boundary of the 2+1D XY model coexist with the LL edge. We have
\begin{eqnarray} S &=& S_{\rm ord} + \frac{1}{2g} \int dx d\tau \,\left(\frac{1}{v_s} (\d_\tau \theta)^2 + v_s (\d_x \theta)^2\right)  \nn\\
&-&
\frac{\tilde s v_b }{2} \int dx d \tau\,  \left( e^{i \theta} \hat{\phi}^* + e^{-i \theta} \hat{\phi}\right). \label{SordLL}\end{eqnarray}
Here $S_{\rm ord}$ is the action of the 2+1D XY model with ordinary BCs and the complex field $\hat{\phi}$ is the corresponding boundary order parameter. $\tilde s$ is the dimensionless coupling constant between $\hat{\phi}$ and the LL XY order parameter $e^{i \theta}$. The dimension of $e^{i \theta}$ is $K = \frac{g}{4\pi}$. A key point that distinguishes the edge of the Kane-Mele-Hubbard model from that of a classical 3D XY model is that phase slips of $e^{i \theta}$ carry electron number and are, therefore, prohibited in the action by the electron number symmetry. The electric charge carried by the spin order parameter phase slips is a direct consequence of the mixed anomaly between the spin and charge symmetries on the edge of a quantum spin Hall insulator and reflects the topological nature of the latter. For the 2+1D XY ordering transitions out of a conventional paramagnet (superfluid) with just $U(1)_{\rm spin}$ ($U(1)_{\rm charge}$) internal symmetry, the absence of phase slips on the boundary can be protected by translational symmetry along the edge if the edge magnetization (charge density) is incommensurate --- such a case was considered in Ref.~\cite{Metlitski-20}.

We now review the RG analysis of the action (\ref{SordLL}). For $K > 2 -\Delta_{\hat{\phi}}$ the interaction $\tilde{s}$ is irrelevant and the ordinary + LL edge phase is stable. As $K$ approaches $2 -\Delta_{\hat{\phi}}$, the coupling constants run as\cite{Metlitski-20}:
\begin{eqnarray}
\frac{d \tilde{s}}{d \ell} &\approx& (2 -\Delta_{\hat{\phi}} - \frac{g}{4\pi}) \tilde{s},\nn\\
\frac{dg}{d \ell} &\approx& -\frac{1}{2} (A+B) \tilde{s}^2 g^2, \\
\frac{d (v_s/v_b)}{d \ell} &\approx& \frac{1}{2} (A-B) g \tilde{s}^2 \frac{v_s}{v_b}
\end{eqnarray}
with 
 \begin{eqnarray}
  A(v_s/v_b) = \frac{v_b}{4 v_s} \int_0^{2\pi} d \theta \frac{\cos^2 \theta}{(\cos^2 \theta + \frac{v^2_s}{v^2_b} \sin^2 \theta)^{\frac{g}{4\pi}}}, \nn\\
 B(v_s/v_b) = \frac{v_s}{4 v_b} \int_0^{2\pi} d \theta \frac{\sin^2 \theta}{(\cos^2 \theta + \frac{v^2_s}{v^2_b} \sin^2 \theta)^{\frac{g}{4\pi}}}.    
 \end{eqnarray} 
It is convenient of make a change of variables
\begin{equation} u = \frac{g}{4\pi} - (2-\Delta_{\hat{\phi}}), \quad v = \sqrt{2 \pi (A+B)}(2-\Delta_{\hat{\phi}}) \tilde{s}. \end{equation}
Then
\begin{eqnarray} \frac{d v}{d \ell} &\approx& - u v, \nn\\
\frac{d u}{d \ell} &\approx& -v^2, \nn\\
\frac{d (v_s/v_b)}{d \ell} &\approx&  \frac{1}{2-\Delta_{\hat{\phi}}} \frac{A-B}{A+B} v^2  \frac{v_s}{v_b}.\end{eqnarray}
To the present order, the flow equation of $u$ and $v$ are exactly the same as for the usual KT transition, further the flow of $v_s$ does not affect the flow of $u$ and $v$. This leads to a ``special" transition out of the ordinary + LL edge phase with a correlation length that diverges as $\xi \sim \exp({\rm const}/\sqrt{g - g_c})$. At the transition $u$ and $v$ flow to zero along the separatrix $u(\ell) = v(\ell) = \frac{v(0)}{1 + v(0) \ell}$ and the scaling dimension of the boundary AFM order parameter is given by $\Delta_{e^{i \theta}} = 2 - \Delta_{\hat{\phi}}\approx 0.77$. Further, the  ratio of edge to bulk velocity remains non-universal. 

We remark in passing that to this order in $\tilde s$ exactly the same RG equations govern the special transition out of the ordinary + LL phase when the bulk AFM order parameter has Ising symmetry, with the appropriate modification of the ordinary boundary order parameter scaling dimension $\hat{\Delta}_\phi$.\cite{JianSK24, ge25}  

For $g$ smaller than the critical value corresponding to the special transition described above we loose analytical control of the action (\ref{SordLL}). It is expected that the theory flows to the extraordinary-log edge phase, which is described by the action
\begin{equation}  S = \frac{1}{2g}  \int dx d \tau  \Big[\frac{1}{v_s}(\partial_\tau \theta)^2 
+ v_s (\partial_x \theta)^2\Big] + \ldots,
\label{Seo}
\end{equation}
where the dots stand for the universal coupling of the edge  AFM order parameter $e^{i\theta}$ to the boundary fields of the 2+1D XY bulk with ``normal" BCs. This coupling leads to the RG running of $g$ and $v_s$ given by\cite{Metlitski-20}
\begin{align}
\frac{dg}{d \ell} &= -\frac{\alpha}{2} \left(\frac{v_s}{v_b} + \frac{v_b}{v_s}\right) g^2, \\ 
\frac{d}{d\ell} (v_s/v_b) &= - \frac{\alpha g}{2} \left[(v_s/v_b)^2 - 1\right],
\end{align}
where $v_b$ is the bulk velocity, which does not run, and $\alpha$ is a universal number estimated to be $\alpha \approx 0.300(5)$\cite{PTM-21}. At the longest length scales $v_s/v_b$ flows to $1$ and  $g$ runs logarithmically to zero as $g(\ell) \sim \frac{1}{\alpha \ell}$.

\section{Spin stiffness in QMC}
Here we will consider a generic Hubbard model, 
\begin{equation}
  \hat{H}   =   \sum_{\vec{\imath},\vec{\jmath}} \hat{c}^{\dagger}_{\vec{\imath}}\ t^{\phantom\dagger}_{\vec{\imath},\vec{\jmath}}\ \hat{c}^{\phantom\ddagger}_{\vec{\jmath}} +  \frac{U}{2}  \sum_{\vec{\imath}}  \left( \hat{c}^{\dagger}_{\vec{\imath}}\hat{c}^{\phantom \dagger}_{\vec{\imath}}  -1 \right)^2 
\end{equation} 
where $\hat{c}^{\dagger}_{\vec{\imath}}   = \left(\hat{c}^{\dagger}_{\vec{\imath},\uparrow}, \hat{c}^{\dagger}_{\vec{\imath},\downarrow}  \right) $   is a two-component spinor  and  $t_{\vec{\imath},\vec{\jmath}}$ a  $2\times 2$ matrix.   To compute the spin  stiffness we  twist the BCs in the x-direction:   
\begin{equation}
  \hat{c}^{\dagger}_{\vec{\imath} + L\vec{e}_x} =  \hat{c}^{\dagger}_{\vec{\imath}}  U(\vec{e},\varphi) 
\end{equation}
with 
$U(\vec{e},\varphi)  = e^{i \varphi\vec{e}\cdot \vec{S}}  $, where 
$\vec{S}$ are the generators of SU(2) with normalization $\text{Tr} \left[ S_{\alpha}S_{\beta}\right] = \frac{\delta_{\alpha,\beta}}{2}$ and $\vec{e}$ is the twist direction in the spin space.    The spin stiffness is then defined  in Eq.~(\ref{rhodef}). 
For best performance  it is convenient to consider a canonical  transformation  that distributes the twist over all x-bonds:  
\begin{equation}
    \hat{f}^{\dagger}_{\vec{\imath}} = \hat{c}^{\dagger}_{\vec{\imath}} U\left(\vec{e},-\frac{i_x}{L}\varphi\right). 
\end{equation}
Under this transformation, the $f$ fermion satisfies periodic BCs in the x-direction and the    Hamiltonian transforms to:
\begin{eqnarray}
  \hat{H}  = & &  \sum_{\vec{\imath},\vec{\jmath}} \hat{f}^{\dagger}_{\vec{\imath}} U\left(\vec{e},\frac{i_x}{L}\varphi\right) t^{\phantom\dagger}_{\vec{\imath},\vec{\jmath}}\ U\left(\vec{e},-\frac{j_x}{L}\varphi\right)\hat{f}^{\phantom\ddagger}_{\vec{\jmath}}   \nonumber \\ 
  & &   + \frac{U}{2}  \sum_{\vec{\imath}}  \left( \hat{f}^{\dagger}_{\vec{\imath}}\hat{f}^{\phantom\dagger}_{\vec{\imath}}  -1 \right)^2   \nonumber \\
   =  & &    \hat{H}_0   +  \hat{H}_1 \frac{\varphi}{L}  + \hat{H}_2  \left(\frac{\varphi}{L} \right)^2 + {\cal O}\left( \frac{\varphi}{L} \right)^3.
\end{eqnarray}
In the previous equation,    $\hat{H}_0$   corresponds to the Hamiltonian without twist,  and 
\begin{eqnarray}
  \hat{H}_1 = & &  \sum_{\vec{\imath},\vec{\jmath}} \hat{f}^{\dagger}_{\vec{\imath}}  \left( 
    i \vec{e} \cdot \vec{S} \, i_x t_{\vec{\imath},\vec{\jmath}} - t_{\vec{\imath},\vec{\jmath}}\ i \vec{e} \cdot \vec{S} \, j_x 
    \right)  \hat{f}^{\phantom\ddagger}_{\vec{j}}    \\
  \hat{H}_2 = & & 
  \sum_{\vec{\imath},\vec{\jmath}} \hat{f}^{\dagger}_{\vec{\imath}}  \left( 
    \vec{e} \cdot \vec{S} \, i_x t_{\vec{\imath},\vec{\jmath}} \, j_x \vec{e} \cdot \vec{S} 
  - \frac{1}{8} (i_x^2 + j_x^2)t_{\vec{\imath},\vec{\jmath}}\right)  \hat{f}^{\phantom\ddagger}_{\vec{\jmath}}   \nonumber 
\end{eqnarray}
Expanding  the logarithm of the partition function $Z(\varphi)$  up to second order in the twist gives: 
\begin{eqnarray}
   & &  \ln Z(\varphi)  =   \ln Z(0) - \left(\frac{\varphi}{L}\right)^2 \beta\times   \\  
   & & \left(  \langle \hat{H}_{2} \rangle     - 
    \int_{0}^{\beta} d \tau \left[ \left< \hat{H}_{1}(\tau) \hat{H}_{1}(0) \right> -  
     \left< \hat{H}_{1} \right>^2  \right] \right) +  {\cal O} \left( \varphi^3  \right)  \nonumber 
\end{eqnarray}
such that the spin stiffness reads:
\begin{equation}
    \rho =  \frac{2}{L^d} \left(  \langle \hat{H}_{2} \rangle     - 
    \int_{0}^{\beta} d \tau \left[ \left< \hat{H}_{1}(\tau) \hat{H}_{1}(0) \right> -  
     \left< \hat{H}_{1} \right>^2  \right] \right). 
\label{eq:stiffness}
\end{equation} 
In our   we consider  $\vec{e} = \hat{e}_z$ since the spin orders in the x-y  plane.  For the U(1) spin symmetric Kane-Mele model that hopping matrix is diagonal in the spin indices.  This leads to  simplifications in the above expressions.  In particular: 
\begin{equation}
    \hat{H}_1 =  \sum_{\vec{\imath},\vec{\jmath}} \hat{f}^{\dagger}_{\vec{\imath}} i \vec{e} \cdot \vec{S} (i_x - j_x)  \hat{f}^{\phantom\ddagger}_{\vec{\jmath}}
\end{equation}
and 
\begin{equation}
  \hat{H}_2 =  - \sum_{\vec{\imath},\vec{\jmath}} \hat{f}^{\dagger}_{\vec{\imath}} \frac{1}{8} t_{\vec{\imath},\vec{\jmath}}\ (i_x - j_x)^2  \hat{f}^{\phantom\ddagger}_{\vec{\jmath}}.
\end{equation}
The expression of Eq.~(\ref{eq:stiffness})   can readily be computed within the  ALF implementation of the auxiliary field QMC algorithm \cite{ALF_v2}.

\section{Bulk and boundary velocities}
\begin{figure}
    \centering
    \includegraphics[width=0.9\linewidth]{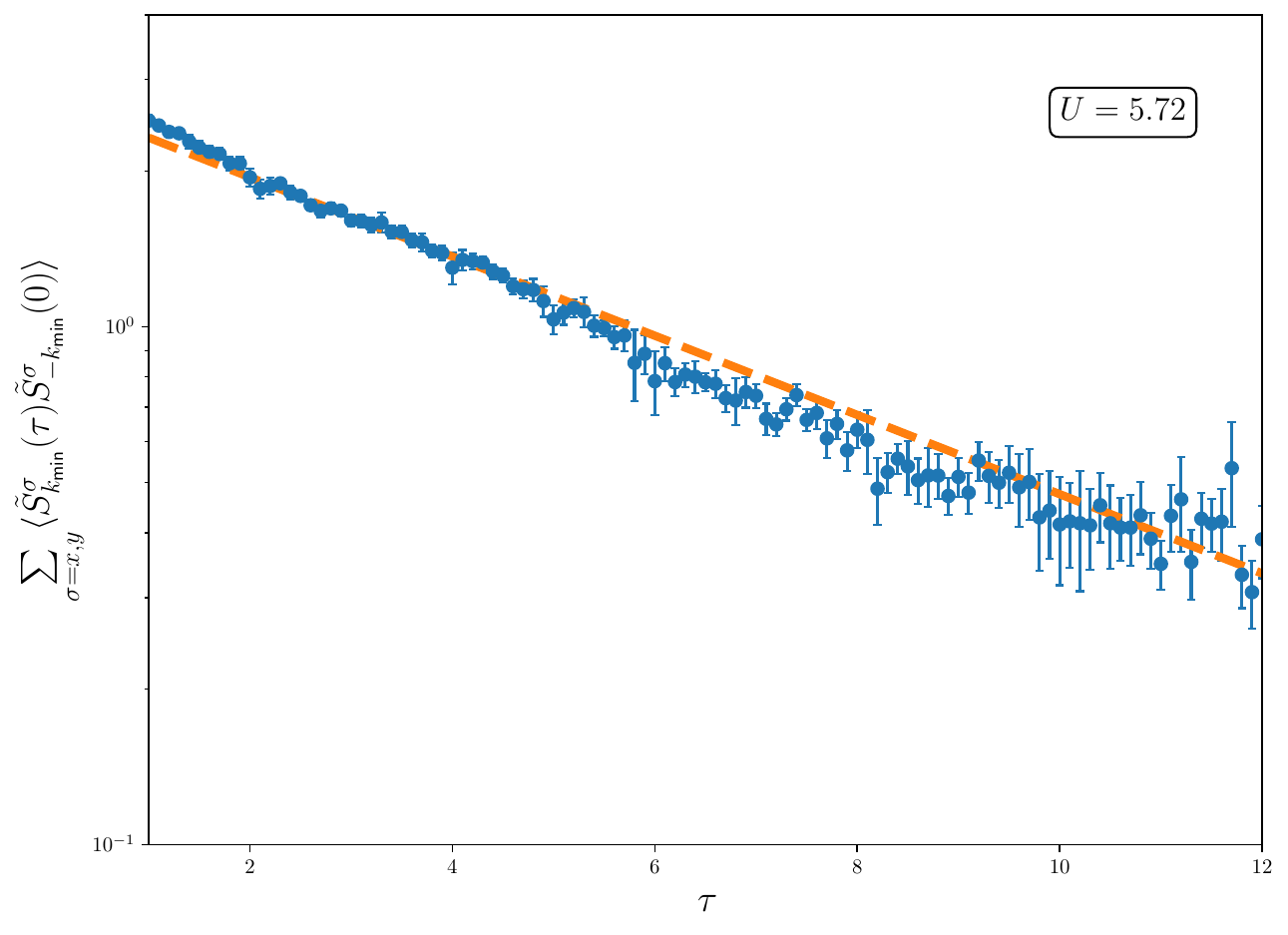}
    \caption{Imaginary-time bulk spin correlations, obtained from simulations with periodic BCs [Hamiltonian (\ref{kmh})] close to the quantum critical point.
    We show the correlations obtained for lattice size $L=28$ and the smallest nonzero momentum $(0, 4\pi/\sqrt{3}/28)$. Dashed line is a fit to Eq.~(\ref{corr_tau_kmin}), obtained employing the data for $2\le \tau \le 4$. }
    \label{fig:corr_tau_kmin_bulk}
\end{figure}
As we discuss in the main text,
the bulk and the boundary velocities enter in the RG flow of the model in the extraordinary-log phase.
Their ratio $v_s/v_b$ is predicted to very slowly flow to 1, as a power of $\ln L$ \cite{Metlitski-20}, such that on feasible range of lattice sizes, one observes an effective $v_s/v_b \approx \text{const} \ne 1$, resulting in a correction to the observed parameter $\alpha$ [See Eq.~(\ref{alphar})].
The velocities can be estimated from expected exponential decay of the imaginary-time spin correlations, computed in Fourier space at the smallest nonzero momentum, which follows from the linear dispersion in a critical energy spectrum
\begin{equation}
    \sum_{\sigma = x,y}\langle \tilde{S}^\sigma_{k_{\text{min}}}(\tau) \tilde{S}^\sigma_{-k_{\text{min}}}(0) \rangle = Ae^{-vk_{\text{min}}\tau},
    \label{corr_tau_kmin}
\end{equation}
with $v=v_b$ in the case of bulk correlations, and $v_s$ for the boundary ones.

Before we proceed, we note several caveats for the above procedure. First, to extract the energy gap above the ground state from the imaginary time correlator, we would ideally be working in the limit $\beta \gg L$. Second, for the case of the bulk velocity, the spectrum of the 2+1D XY model on the torus is not that of a free theory and there are corrections to the relation $E_{\vec{k}} - E_0 = v_b |\vec{k}|$. Nevertheless, based on $\epsilon$-expansion and  exact diagonalization studies the corrections are numerically rather small and will be ignored in our analysis.\cite{WhitsittTorus} 

In Fig.~\ref{fig:corr_tau_kmin_bulk} we show the bulk correlations, for the largest available lattice $L=28$ and at the smallest momentum $(0, 4\pi/\sqrt{3}/28)$.
QMC data show a qualitative exponential decay, albeit with increasing fluctuations for $\tau \gtrsim 4$, reflecting well known numerical instabilities \cite{AF_notes}.
We fit the QMC to the right-hand side of Eq.~(\ref{corr_tau_kmin}), employing only a window of data at $2\le\tau\le 4$.
This choice is somewhat arbitrary, but it is motivated by the need to avoid nonuniversal decay at small $\tau$ and noisy data at larger $\tau$.
The fit delivers an estimate $v_b = 0.68(5)$.
In Fig.~\ref{fig:corr_tau_kmin_bulk} we also show the fitted curve, which on an optical scale matches well the data.
This analysis neglects corrections to leading behavior in Eq.~(\ref{corr_tau_kmin}), hence the quoted precision should be taken with some grain of salt.
Nevertheless, repeating the analysis for the smallest momentum in the other direction $k_{\text{min}}=(4\pi/28,0)$ gives a perfectly consistent estimate $v_b=0.70(5)$, giving us confidence in the reliability of the result.

\begin{figure}
    \centering
    \includegraphics[width=0.9\linewidth]{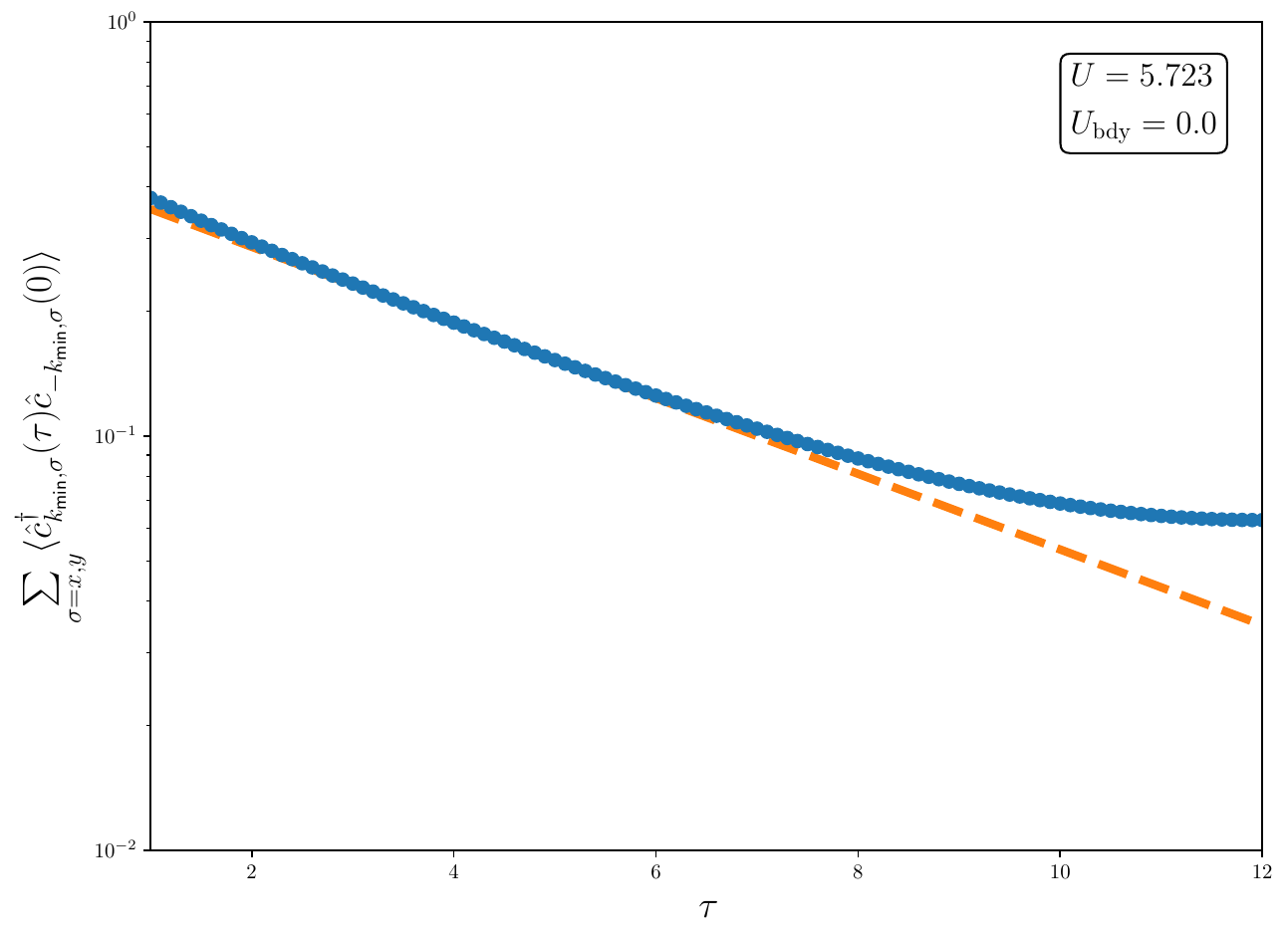}
    \caption{Imaginary-time edge fermion Green's function, obtained from simulations of the zigzag Hamiltonian (\ref{kmh_zigzag}) at the quantum critical point $U=5.723$ and in the ordinary phase $U_{\text{bdy}}=0$.
    We show the correlations obtained for lattice size $L=24$ and the smallest momentum $\pi+2\pi/24$. Dashed line is a fit to Eq.~(\ref{corr_tau_green_kmin}), obtained employing the data for $3.5\le \tau \le 5$. }
    \label{fig:corr_tau_green_kmin_edge}
\end{figure}

To extract the LL velocity in the ordinary phase we use imaginary-time fermion Green's function along the edge, computed at the smallest momentum $k_{\text{min}} = \pi + \frac{2\pi}{L}$, which we fit to
\begin{equation}
    \sum_{\sigma = x,y}\langle \hat{c}^\dag_{k_{\text{min}},\sigma}(\tau)
    \hat{c}^{\phantom{\dag}}_{k_{\text{min}},\sigma}(0)\rangle = 
    Ae^{-v_s\frac{2\pi}{L}\tau},
    \label{corr_tau_green_kmin}
\end{equation}
In Fig.~\ref{fig:corr_tau_green_kmin_edge} we show the Green's function for our largest lattice $L=24$, together with a fit of QMC data for $3.5\le \tau\le 5.5$.
This choice gives a reasonably good $\chidof$, and delivers $v_s =0.804(2)$.
As for the determination of the bulk velocity, the error bar should be taken with some grain of salt.
On varying the interval of fitting, we obtain a consistent value $v_s\approx 0.8$.

\begin{figure}[t]
    \centering
    \includegraphics[width=0.9\linewidth]{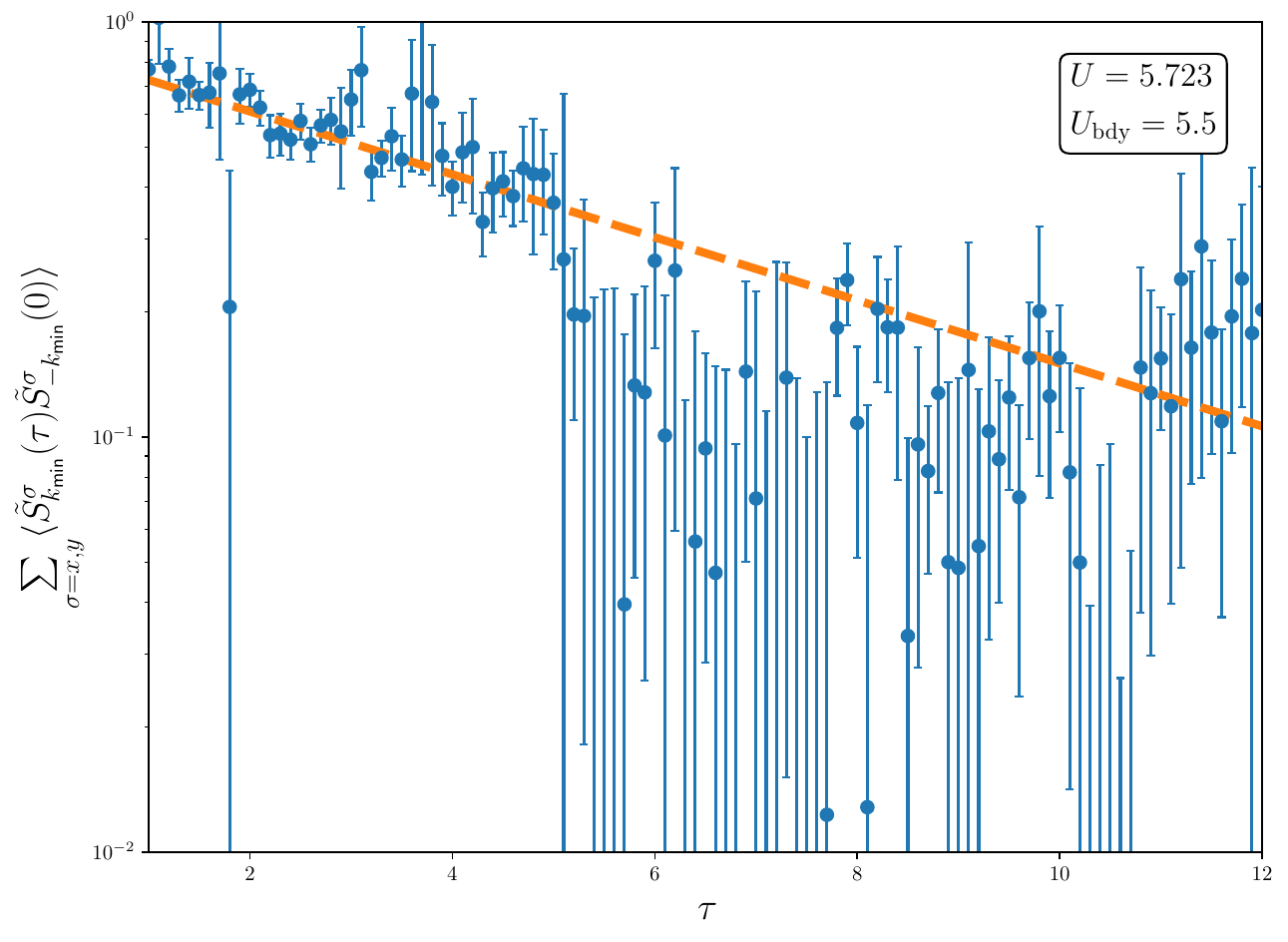}
    \caption{Imaginary-time edge spin correlations, obtained from simulations of the zigzag Hamiltonian (\ref{kmh_zigzag}) at the quantum critical point $U=5.723$ and in the extraordinary-log phase $U_{\text{bdy}}=5.5$.
    We show the correlations obtained for lattice size $L=24$ and the smallest nonzero momentum $2\pi/24$. Dashed line is a fit to Eq.~(\ref{corr_tau_kmin}), obtained employing the data for $2\le \tau \le 4$. }
    \label{fig:corr_tau_kmin_edge}
\end{figure}
In the main text, we have extracted the boundary velocity $v_s$ in the extraordinary-log phase from the fits of the  susceptibility and the stiffness [Eqs.~(\ref{fit_chi})-(\ref{valpharDB})].
As a check, we have also estimated $v_s$ using Eq.~(\ref{corr_tau_kmin}).
In Fig.~(\ref{fig:corr_tau_kmin_edge}) we show the spin correlations on the edge, for the largest available lattice $L=24$, in the extraordinary-log phase.
The QMC data are affected by considerable numerical instabilities, such that only a rough estimate of $v_s$ can  be extracted.
Employing in the fit the data for $2\le\tau\le 4$, where QMC data are comparably more stable, delivers the estimate $v_s=0.67(20)$, in agreement with the value $v_s=0.9(1)$ quoted in the main text.

\begin{comment}

\section{Single particle Green functions in the helical liquid}
Some care has to be taken when computing the single particle Green functions in the helical liquid phase.  To illustrate the point, we model the helical liquid with 
\begin{equation}
  \hat{H} = v_f \sum_{k} \hat{c}^{\dagger}_{k} k \sigma^z \hat{c}^{\phantom\dagger}_{k} 
\end{equation}
with $\hat{c}^{\dagger}_{k}$ a two-component spinor. 
For the above Hamiltonian and in the  zero temperature limit, 
\begin{equation}
  \left< \hat{c}^{\dagger}_{r, \uparrow} \hat{c}^{\phantom\dagger}_{0,\uparrow} \right>  = 
  \frac{1}{2\pi} \frac{1}{ir} \left( 1 - e^{-i \Lambda r}\right)
\end{equation}
where $\Lambda $  is a short distance cutoff. The down Green function is obtain by time reversal symmetry.  Hence to at best compute  extract the scaling dimention of the fermion -- $1/2$ in this non-interacting case -- it is convenient to consider 
\begin{equation}
\left| \sum_{\sigma} \sigma \left< \hat{c}^{\dagger}_{r,\sigma} \hat{c}^{\phantom\dagger}_{0,\sigma} \right> \right| = \frac{1}{\pi r}. 
\end{equation}
No such care has to be taken when computing the spectral function.  In particular: 
\begin{equation}
   A(k\omega) =  - \frac{1}{\pi} \text{Im} \text{Tr} 
   G^{ret}(k,\omega) 
\end{equation}
with 
\begin{equation}
   G^{ret}(k,\omega) =  \frac{1}{\omega + i0^{+} - v_f \sigma^z k}
\end{equation}
gives $A(k,\omega) = \sum_{\sigma} \delta(\omega - v_f \sigma k)$, as expected.  This equally corresponds to the quantity we compute in the main text.
\end{comment}

\section{Tables of fits}
In Table \ref{tab:fits_bulk} we show fits of $\xi/L$ to the right-hand side of Eq.~(\ref{xil_scaling_taylor}), as a function of the minimum lattice size \Lmin taken into account.
\begin{table*}
    \centering
    \caption{Fits of $\xi/L$ for the bulk phase transition to Eq.~(\ref{xil_scaling_taylor}) for $m=1$, $k=0$, and as a function of the minimum lattice size \Lmin taken into account.}
    \begin{ruledtabular}
    \begin{tabular}{Lwwwwq}
    \multicolumn{1}{c}{\Lmin} & \multicolumn{1}{c}{$U_c$} & \multicolumn{1}{c}{$f_0$} & \multicolumn{1}{c}{$f_1$} & \multicolumn{1}{c}{$g_0$} & \multicolumn{1}{c}{\chidof} \\
    \hline
     8  & 5.72316(89) &  0.3010(19) &  0.011930(93) &  -0.2451(97) &   0.8 \\
    12  & 5.7224(15)  & 0.2992(39)  & 0.011944(94)  & -0.231(24)  & 0.8 \\
    16  & 5.7234(29)  & 0.3024(91)  & 0.01198(10)   & -0.250(62)  & 0.7 \\
    20  & 5.7225(89)  & 0.299(34)   & 0.01202(14)   & -0.22(26)   & 1.0 \\
    \end{tabular}
    \end{ruledtabular}
    \label{tab:fits_bulk}
\end{table*}

\begin{table*}
    \centering
    \caption{Fits of $\chi/L$ for Eq.~(\ref{fit_chi}) and of $\rho L$ to Eq.~(\ref{fit_rho}) in the extraordinary-log phase, as a function of the minimum lattice size \Lmin taken into account.
    For each fit we compute the edge velocity $v_s$ and the coefficient $\alpha_r$, computed with Eq.~(\ref{valpharDB}).
    }
    \begin{ruledtabular}
    \begin{tabular}{Lwwq|wwq|ww}
    \multicolumn{1}{c}{\Lmin} & \multicolumn{1}{c}{$A$} & \multicolumn{1}{c}{$B$} & \multicolumn{1}{c}{\chidof} & \multicolumn{1}{c}{$C$} & \multicolumn{1}{c}{$D$} & \multicolumn{1}{c}{\chidof} &  \multicolumn{1}{c}{$v_s$} & \multicolumn{1}{c}{$\alpha_r$} \\
    \hline
       8 &  0.483(42) &  0.412(14) & 2.00   &  -0.300(36) &  0.328(14) & 0.34  &   0.892(25) & 0.183678(71) \\
      12 &  0.383(65) &  0.445(22) & 1.02   &  -0.316(96) &  0.333(33) & 0.49  &   0.866(48) & 0.19244(34) \\
      16 &   0.29(12) &  0.476(38) & 1.00   &  -0.43(22)  &  0.370(74) & 0.66  &   0.882(95) & 0.2099(16) \\
    \end{tabular}
    \end{ruledtabular}
    \label{tab:fits_chirho}
\end{table*}
In Table \ref{tab:fits_chirho} we report fits of $\chi/L$ for Eq.~(\ref{fit_chi}) and of $\rho L$ to Eq.~(\ref{fit_rho}), as a function of the minimum lattice size \Lmin taken into account.

\end{document}